\documentclass[useAMS,usenatbib,usegraphicx]{mn2e}

\usepackage{psfig}
\usepackage{times}
\usepackage{subfigure}

\newcommand{\kms}{\hbox{km s$^{-1}$}}

\newcommand{\vsini}{\hbox{$v$\,sin\,$i$}}

\newcommand{\ha}{H$_{\alpha}$}
\newcommand{\hb}{H$_{\beta}$}
\newcommand{\hg}{H$_{\gamma}$}
\newcommand{\hd}{H$_{\delta}$}
\newcommand{\he}{H$_{\epsilon}$}
\newcommand{\height}{H$_{8}$}
\newcommand{\hnine}{H$_{9}$}
\newcommand{\hten}{H$_{10}$}

\newcommand{\speedy}{\hbox{Speedy Mic}}

\begin{document}

\title[Speedy Mic prominence system - II]{The coronal structure of Speedy Mic - II: Prominence masses and off-disc emission}

\makeatletter
 
\def\newauthor{%
  \end{author@tabular}\par
  \begin{author@tabular}[t]{@{}l@{}}}
\makeatother

\author[N.J.~Dunstone, A.~Collier~Cameron, J.R.~Barnes, M.~Jardine]
{N.J.~Dunstone$^1$\thanks{E-mail: njd2@st-andrews.ac.uk} A. Collier Cameron$^1$ J.R. Barnes$^2$ M. Jardine$^1$\\
$^1$ SUPA\thanks{Scottish Universities Physics Alliance},School of Physics and Astronomy, University of St Andrews, Fife KY16 9SS. UK. \\
$^2$ Centre for Astrophysics Research, University of Hertfordshire, College Lane, Hatfield, Herts AL10 9AB. UK.}

\date{2006, 200?}

\maketitle

\begin{abstract}
Observations of stellar prominences on young rapidly rotating stars provide unique probes of their magnetic fields out to many stellar radii.  We compare two independently obtained datasets of the K3 dwarf \speedy\ (BO Mic, HD 197890) using the Anglo-Australian Telescope (AAT) and the ESO Very Large Telescope (VLT). Taken more than a fortnight apart they provide the first insight into the evolution of the prominence system on such a young rapidly rotating star.  The largest prominences observed transiting the stellar disc are found at very similar rotational phases between the epochs.  This suggests that the magnetic structures supporting the prominences retain their identity on a two to three week timescale.  By taking advantage of the high signal-to-noise and large wavelength range of the VLT observations we identify prominences as transient absorption features in all lines of the Hydrogen Balmer series down to \hten.  We use the ratios of the prominence EWs in these lines to determine their column densities in the first excited state of hydrogen.  We determine the optical depths, finding prominences to be rather optically thick ($\tau\approx20$) in the \ha\ line.  The total hydrogen column density and thus the prominence masses are determined via observations of the CaII H\&K lines.  We find typical masses for four of the largest prominences to be {{in the range $0.5 - 2.3\ \mathrm{x}10^{14}\ \mathrm{kg}$, slightly larger than giant solar prominence masses.}} Rotationally modulated emission is seen outside of the \ha\ line.  These loops of emission are shown to be caused by prominences seen off the stellar disc.  We find that all of the large emission loops can be associated with prominences we see transiting the stellar disc.  {{This combined with the fact that many prominences appear to eclipse the off-disc emission of others, strongly suggests that the prominence system is highly flattened and likely confined to low stellar latitudes.}}

\end{abstract}


\section{INTRODUCTION}
\protect\label{sect:intro}

Solar prominences are either observed against the surface of the Sun as dark absorption structures known as `filaments' or off the solar limb as bright emission loops.  Traditionally to observe such features a narrow-band filter is used centred on 6563\AA.  This corresponds to the first and strongest transition of the hydrogen Balmer series, hydrogen alpha (\ha). 

Stellar prominences were first reported by \cite{cam1989a} as rapid transient absorption features passing through the Doppler broadened \ha\ line of the rapidly rotating K0 dwarf AB Doradus (AB Dor).  From subsequent observations these clouds of material were shown to be in {{presumably magnetically enforced}} co-rotation with the stellar surface and so dubbed ``prominences''. Since their discovery prominences have now been reported on around a dozen rapidly rotating late-type stars (see \citealt{cam2003rev} for a recent review).   These observations have been primarily restricted to the strong Balmer lines, often only (\ha) and occasionally hydrogen beta (\hb).  

The AB Dor prominence system remains the most studied to date.  Prominences have consistently been reported (\citealt{cam1989a}; \citealt{cam1989b}; \citealt{donati97ab}; \citealt{cam1999}; \citealt{donati99ab}) to lie at a range of distances from the stellar rotation axis from 2 to 8 $\mathrm{R_*}$, with a concentration at the co-rotation radius (2.7\ $\mathrm{R_*}$ on AB Dor).   

\cite{dunstone06} (hereafter Paper I) presented observations of the prominence system of another rapidly rotating star, Speedy Mic (BO Mic, HD 197890).  Speedy Mic is a K3V dwarf with a $0.38\ \mathrm{d}$ rotational period, an equatorial rotation velocity of \hbox{\vsini\ $\simeq$ 132 \kms} and is seen at a high inclination{{.  The stellar inclination was found to be  $70\pm5\degr$ by \cite{barnes05}, however at such high inclinations Doppler imaging provides poor constraints.  \cite{wolter05a} also found an inclination of $70\degr$, yet they concede that varying the inclination by up to $\pm20\degr$ has little influence on the resulting Doppler maps.}}   Recent studies of X-ray variability have shown Speedy Mic to be one of the most active stars in the solar neighbourhood. \cite{singh99xray} found $\mathrm{log}({L_X}/{L_{bol}})\simeq-3$ placing it in the saturated regime.  In Paper I we found a densely packed co-rotating prominence system, with heights of prominences grouped around 3 $\mathrm{R_{*}}$ from the stellar rotation axis.  This meant that they were not concentrated at the co-rotation radius (1.95 $\mathrm{R_*}$ on \speedy) but in fact at twice the height above the stellar surface.  Clearly such observations of co-rotating prominences so far above the co-rotation radius require them to be supported by closed magnetic loops.  Then prominences should be direct tracers of the stellar magnetic field on the largest scales, out to many stellar radii in the corona.  

X-ray and Zeeman-Doppler Imaging (ZDI) observations are at odds with the hypothesis that large magnetic loops extend high into the corona of such rapidly rotating stars. Recently \cite{hussain_chandra1_05} showed that the closed field containing compact hot X-ray emitting material on AB Dor was close to the stellar surface and unlikely to extend beyond a radius of 1.75 $\mathrm{R_*}$.  From ZDI observations of the surface magnetic field of AB Dor the structure of the X-ray emission was determined assuming a simple heating law (\citealt*{jardine02structure}; \citealt{jardine02xray}).  They found that the X-ray emission must be confined fairly close to the stellar surface.  As \speedy\ is a faster rotator than AB Dor one would expect the coronal extent to be smaller.  These observations are in apparent conflict with the requirement for magnetic support of prominences at heights of several stellar radii.

A model for prominence support which does not require closed magnetic field lines to extend to several stellar radii above the surface was put forward by \cite{jardine05}.  Instead they propose that prominences are supported by the re-connection of oppositely directed wind-bearing field lines over coronal helmet streamers.  This would result in long thin, highly curved, loops extending out to large radii.  These could then fill up with gas that was originally part of the wind, forming the observed prominences. However while \cite{jardine05} show that such magnetic loops can achieve equilibrium they do not attempt to model the prominence structures themselves. In order to test such models we need more information on the physical properties, evolutionary scenarios and lifetimes of prominence systems. 

{{Solar prominences are also be observed in the \ha\ line as emission features seen off the solar limb. Photons at this wavelength originating from the surface of the Sun can be absorbed by a hydrogen atom in the prominence, thus exciting its electron from the first (n=2) to the second (n=3) excited state.  Providing such an atom does not undergo a collision, the electron will de-excite either to the ground state (n=1) via the emission of a Lyman beta (Ly$_{\beta}$) photon or, almost equally likely, back to the first excited state via the release of another \ha\ photon.  Ly$_{\beta}$ photons are re-absorbed almost immediately, creating new \ha\ photons.  As this emission is isotropic most photons originating from the part of the solar surface occulted by a prominence are `scattered' out of our line-of-sight, creating an absorption feature.  Similarly, photons can be scattered into our line-of-sight by a prominence seen off the solar limb, creating an emission feature. Observing a stellar prominence as both an absorption and an emission feature could potentially provide important insights regarding its physical and geometrical properties.}}

The initial motivation for this study arose from a fortuitous coincidence.  Two weeks after the AAT \speedy\ observations used in Paper I, a separate team performed very similar high-resolution spectroscopic observations of \speedy\ with the VLT.  Two papers using this dataset have been published by the original observers, \cite{wolter05a} and \cite{wolter05b}.  \cite{wolter05a} provides a full account of the observations and then goes on to use the red-arm for Doppler imaging.  \cite{wolter05b} focus on the potential of the CaII H \& K lines for future Doppler imaging of the stellar chromosphere.

Here we compare the two sets of observations in \S \ref{sect:obs},  then in \S \ref{sect:masses} we use the VLT observations to obtain prominence physical properties.  In \S \ref{sect:emiss} we report the observation of loops of emission and show that these are caused by prominences seen off the stellar disc.   We discuss our results in the context of the coronal structure of \speedy\ in \S \ref{sect:discuss}.  Finally we present our conclusions in \S \ref{sect:conc}.

\section{OBSERVATIONS}
\protect\label{sect:obs}

The Speedy Mic observations were obtained from the ESO Science Archive Facility.  They consist of two nights (2002 August 02 and 07) of data using the 8-m Very Large Telescope (VLT) UT 2.  The spectra cover a large wavelength range by using the ${\mathrm{UVES}}$ spectrograph in dichroic mode, 3260\AA\ to 4450\AA\ (the ``blue-arm'') and 4760\AA\ to 6840\AA\ (the ``red-arm'').  A 1\arcsec slit width gave a spectral resolution of ${\lambda}/\Delta{\lambda}\approx40\ 000$. Speedy Mic was observed continuously on both nights yielding complete and uninterrupted phase coverage.

We re-extracted the data using {\sc{ECHOMOP}}, the \'{e}chelle reduction package developed by \cite{mills92} and its implementation of the optimal extraction algorithm developed by \cite{horne86extopt}.  During the data reduction we encountered the same problem reported by \cite{wolter05a} of variations in the MIT-CCD red-arm continuum, {{presumably due to the CCD manufacturing process}}.  In order to correct for the introduced small scale structure we created a ``balance frame'' that included these variations.  

In total there are 125 spectra on August 02, each with an exposure time of 200 s.  The readout time of 70 s makes the cadence of exposures 270 s, therefore 122 exposures were required to cover the 0.38 day period of \speedy.  The situation on August 07 is slightly more complicated with the exposure times being adjusted between 120 s and 200 s to compensate for variable sky conditions.  Four spectra had to be abandoned during reduction due to over-exposure, yielding a total of 141 usable spectra.

After each spectrum was extracted we normalised the spectral orders by fitting the continuum using splines.  Due to the difficulty in locating the continuum level in the blue orders we used a two-dimensional fit that constrained the local shape of the continuum.  We discuss the limitations of this technique in \S \ref{sect:masses}.  The resulting signal-to-noise (S/N) of individual spectra were very high.  For example near the centre of the order containing the \ha\ line the unbinned S/N ranges between 350 and 600 per 0.05\AA\ wavelength element throughout the night.

\subsection{A fortnight of evolution}
\protect\label{sect:evol}

Thanks to the fortuitous coincidence of two different Doppler imaging groups (using different telescopes) both observing \speedy\ within two weeks of each other we now have an exciting and unique opportunity.  The two to three week time gap is not normally an easy one to achieve with high-resolution spectroscopy.  Practical considerations like instrument availability on any given telescope normally limit the time span of observations at any given epoch to mere days.

\begin{figure*}
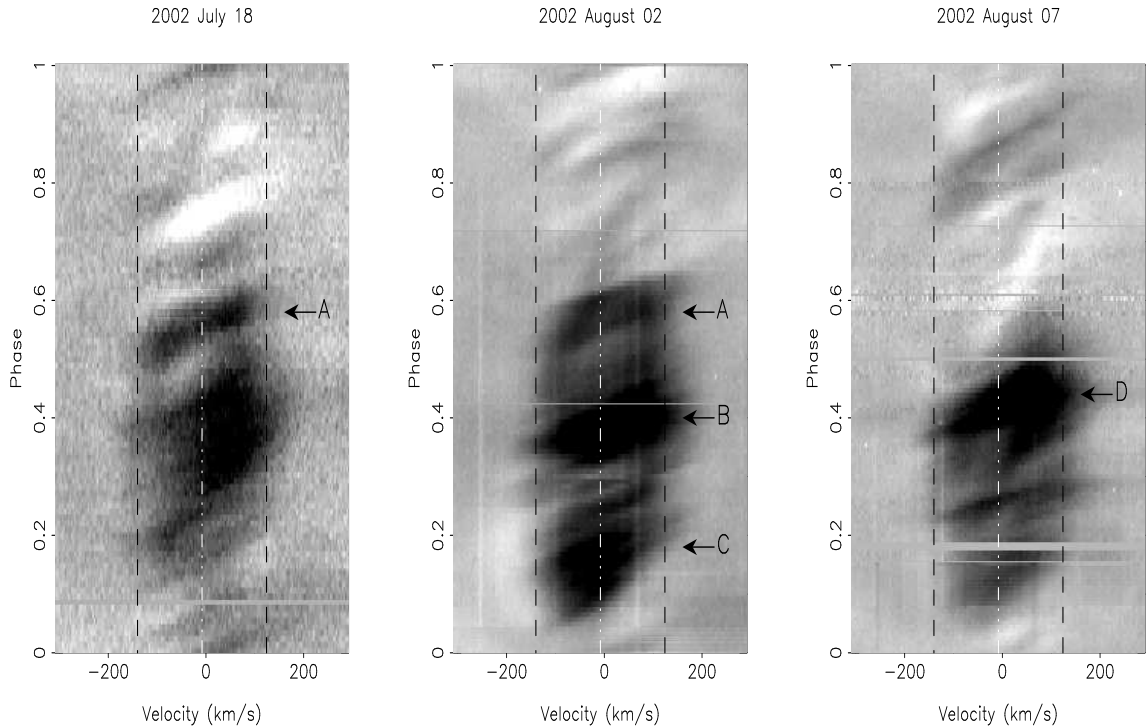

\begin{center}

  \begin{tabular}{ccc}
    \includegraphics[width=5.0cm,height=10cm,angle=0]{Image/njd_speedy06_fig1.ps} &
    \hspace{-2mm}
    \includegraphics[width=5.0cm,height=10cm,angle=0]{Image/njd_speedy06_fig2.ps} &
     \hspace{-2mm}
    \includegraphics[width=5.0cm,height=10cm,angle=0]{Image/njd_speedy06_fig3.ps} \\
  \end{tabular}

\end{center}
\caption[Speedy Mic \ha\ time series spectra ]{Raw \ha\ timeseries spectra for 2002 July 18, August 02 \& 07, with phase plotted against velocity.  Dashed black lines show the \vsini\ limits and the central dashed white line shows the radial velocity of Speedy Mic.  The grey-scale runs from black at 0.85 times the local continuum level, to white at 1.05 times continuum.  The large prominences discussed in the text are labelled.}
\protect\label{fig:rtseries}
\end{figure*}

In Fig.\ref{fig:rtseries} we have plotted the \ha\ trailed spectrum (timeseries) of July 18 from Paper I and compare it to August 02 and 07.  Note that for comparison with this work the July 18 timeseries has been re-phased taking the epoch of phase zero to be the first observation on August 02.   There are fifteen days between July 18 and August 02, that equates to approximately 40 stellar rotations and there are twenty days or 53 rotations between July 18 and August 07. Despite the differences in the quality of the data between the two epochs it is clear to see that qualitatively the timeseries appear similar.

  A visual examination of Fig.\ref{fig:rtseries} shows that August 02 does indeed share features of both July 18 and August 07.  We can see that one half (the `dark' side) of the \speedy\ rotation still has increased absorption due to large prominences.  In the other half (the `light' side) of the stellar rotation cycle the surface is more easily seen and we observe weaker prominences embedded in and around bright surface `plage' features.  Identifying and comparing individual prominences is difficult as there has been considerable evolution between the July and August epochs. Very few of the weaker prominences observed on the light side can be re-identified.  The large dark region at $\phi\approx0.35$ appears stronger and more compact in phase in the August observations. The strong group of prominences at $\phi\approx0.15$ seen on both August epochs has no counterpart in the July observations.

The only prominence that can be compared with any confidence is prominence A at $\phi\approx0.55$.  This prominence was reported in Paper I (again labelled `A') as being present on all three nights (July 18, 19 and 23).  On August 02 we find a prominence at a very similar phase.  Yet on August 07 there is no evidence of the prominence - only a bright surface region.  So can this prominence seen on August 02 really be the same structure that was observed fifteen days earlier?  The main properties of a prominence that we can measure at {\em{both}} epochs are its location in the rotational phase, the projected distance from the \speedy\ rotation axis (height) of the prominence and the maximum fractional depth below the continuum.  The first two of these are found from the Gaussian matched-filter tracking technique described in detail in Paper I.  The difference in the two phases of observation between July 18 and August 02 is $\Delta\phi=0.007$, which is the same as the phase separation between consecutive exposures.  The height of prominence A was quoted as $2.79\pm0.11$  for July 18 and is found to be $2.99\pm0.07$ on August 02.  The maximum fractional depth relative to the continuum when the prominence is transiting the disc center is $\approx0.87$ on July 18 and $\approx0.88$ on August 02.  

The similarity of the properties of this prominence at both epochs lead us to conclude that this is almost certainly the same structure.  As we shall discuss in \S \ref{sect:discuss}, however, we are not suggesting that this prominence necessarily contains the same material at both epochs.

\subsection{Spectral lines showing prominence signatures}

After each spectral order was extracted we stacked the exposures to create a timeseries for each order.  We then collapsed this timeseries to create a mean spectrum and subtracted it from each individual spectrum.  This left us with only time varying features, i.e. star-spots, surface plage regions and prominences.

We performed a visual inspection of all orders for evidence of prominence absorption features moving rapidly through the line profile, just as we observe in the \ha\ line.  This dataset is ideal for an unbiased search, as it is of exceptional S/N and the large wavelength range gives us plenty of lines to examine.  
As was expected, the lines of the Hydrogen Balmer series all show evidence of prominence material.  Starting with the strong \ha\ line the same characteristic prominence pattern is repeated, with different strengths, in each line.  It is possible to visually identify this pattern all the way down to $H_{14}$ at ${\lambda}=3721$\AA.  Prominences in the later hydrogen lines are so optically thin that only the very strongest prominence is identifiable.  Furthermore the star-spot residuals from the many photospheric metal lines present at these blue wavelengths have strengths of the same order of magnitude as the prominence signatures.  Therefore for the rest of this paper we restrict our analysis to lines down to $H_{10}$ at ${\lambda}=3835$\AA, all of which are shown in Fig.\ref{fig:hall}. 

The singly ionised calcium (CaII) H \& K lines at $\lambda=3968$\AA\ and $\lambda=3933$\AA\ have strong emission cores and show strong prominence absorption signatures.  These lines are discussed at length in \S \ref{sect:caii}.  Finally, we find very weak prominence signatures in the sodium doublet at $\lambda=5890$\AA.  Unfortunately these are poorly placed at the very start of an order; this combined with their strengths make them unsuitable for further analysis.  These were the only lines that we can say with confidence show any prominence material.  

\section{Prominence masses}
\protect\label{sect:masses}

\subsection{Column density of hydrogen in the n=2 level}
\protect\label{sect:hcol}

Due to the large wavelength range and the high S/N of these observations it is possible to compare the strengths of individual prominences in different hydrogen Balmer lines.  In order for such a comparison to be possible each line has been re-binned to the same velocity scale, 3.1 \kms\ per pixel.

The trailed spectra of all eight Balmer lines being considered are shown in Fig. \ref{fig:hall} for August 02. Some of the lines have small problems with their reduction.  \hb\ clearly has four bad spectra, which arise due to a bad column {{in the earlier part of this spectral order.  When {\sc{ECHOMOP}} traces this order it rejects and interpolates over these bad pixels, however this procedure was unsuccessful for these four spectra and a good trace was not obtained. Consequently they are not used in further analysis.}} \hten\ shows a vertical line which is due to another chip defect.  Finally it should be noted that the \he\ line is superimposed with the CaII H line, due to the strong prominences seen in CaII we do not use \he\ further in this analysis.

\begin{figure*}
\begin{center}
\includegraphics[height=22cm,angle=0]{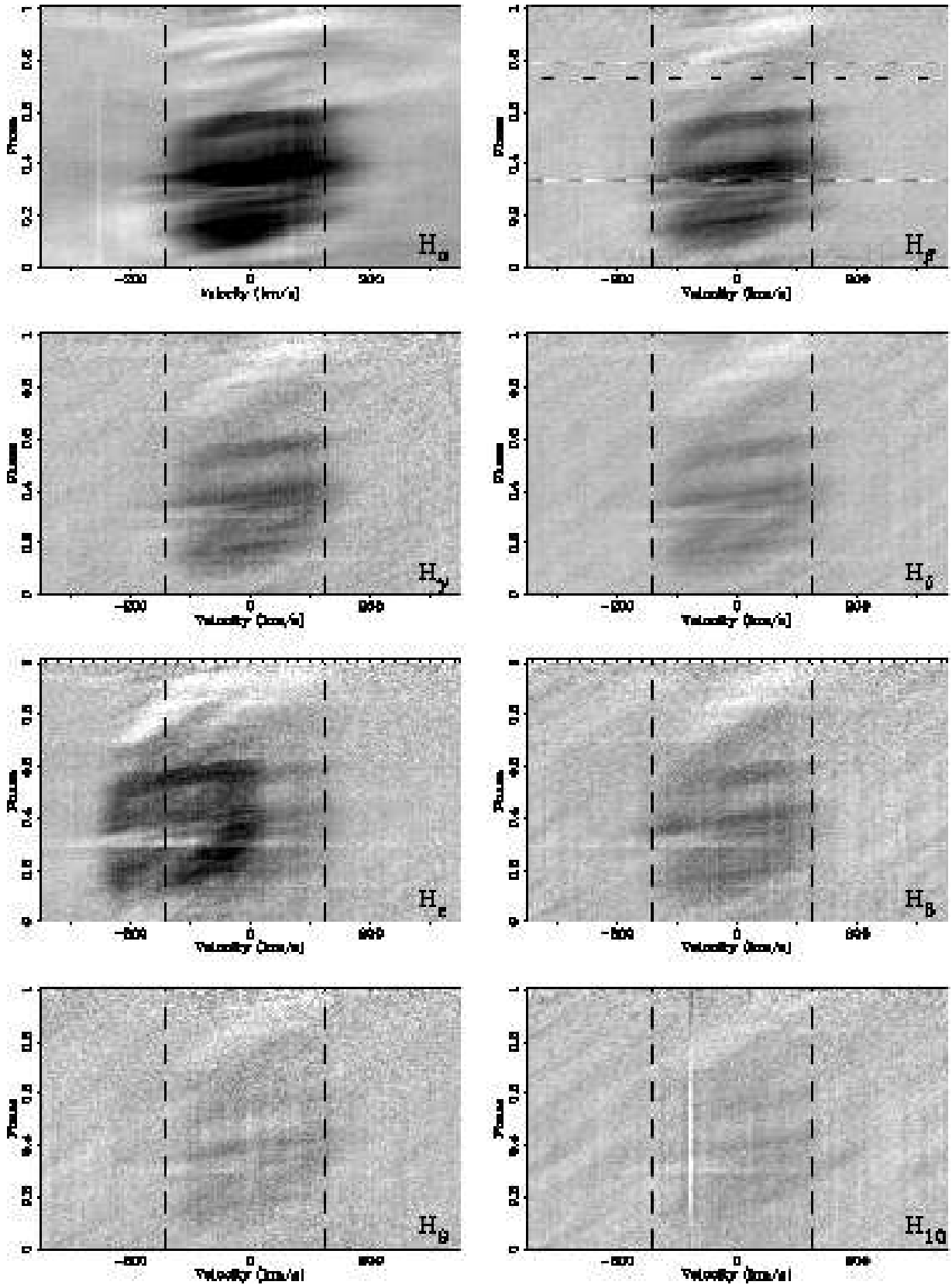}
\end{center}
\caption{The first eight lines of the hydrogen Balmer series for August 02 are shown. {{An average profile has been subtracted from each timeseries (see text).}}  The contrast is such that black is 0.85 and white is 1.05 times the local continuum level.  {{Dashed vertical lines show the \vsini\ limits}}. Note $H_{\epsilon}$ is blended with the {\it{CaII H}} line and $H_{\beta}$ has four bad spectra (see text).}
\protect\label{fig:hall}
\end{figure*}

\subsubsection{Calculating the curve of growth}
\protect\label{sect:curves}

The ratio of the equivalent widths (EWs), or strengths, of individual prominences in different hydrogen lines can be used to calculate the column density of hydrogen atoms in the $n=2$ level.

In order to do this we need to first calculate the absorption cross-section of a hydrogen atom at the centre of each Balmer line.  This is given by:
$$\sigma_{\nu}=\frac{g_u}{g_l}\frac{c^{2}}{8\pi\nu^{2}}\Phi\left(\nu\right)A_{ul}$$
where $g_u$ and $g_l$ are the statistical weights of the upper and lower levels of the transition, $c$ is the speed of light and $\nu$ is frequency.  $A_{ul}$ is the Einstein $A$-coefficient and gives the probability of spontaneous decay from the upper to the lower level.   $\Phi$ is the line profile for which we use a Voigt profile approximation. {{The Lorentz parameter was taken to be natural (radiative) dampening. This is justified because of the scattering mechanism that causes us to observe the prominences in absorption whilst transiting the stellar disc. This tells us that the time between collisions within the prominence is longer than the spontaneous de-excitation time and so collisional broadening is unimportant. We assume the dominant broadening comes from two sources. Firstly the thermal Doppler velocity of hydrogen at a prominence temperature of 10,000 K which is 12.9 \kms\ and secondly a random turbulent velocity.

Solar prominences exhibit random turbulent motions at least as great as a few \kms\ \citep{engvold1998}.  The amount of turbulence we include in the line profile will have a greater effect on the CaII H\&K observations (discussed in \S \ref{sect:caii}) than on those of the hydrogen Balmer lines.  This is because the calcium atom is heavier than the hydrogen atom and so has a much smaller thermal Doppler velocity of 2.0 \kms. We can place an upper limit on the possible turbulent velocity in \speedy\ prominences by measuring the widths of the smallest prominences in the CaII H\&K lines.  A width of approximately 18 \kms\ is found for the smallest absorption features as they cross the centre of the stellar disc. If we assume that these small prominences are unresolved then the measured width should be due to turbulence, but we also need to consider the rotational smearing that occurs during a 200 s exposure. The average prominence height found in Paper I of 3 $\mathrm{R_*}$ corresponds to an orbital velocity of 400 \kms. The \speedy\ radius was calculated in Paper I from a simple dynamical argument to be $1.06\pm{0.04}\ \mathrm{R}_{\odot}$, considerably more bloated {{than}} its K3V main-sequence radius. Therefore during a single exposure such a prominence travels approximately 11\% of the \speedy\ radius which in velocity space is 14 \kms, the majority of the observed width. If we assume that these velocities add in quadrature then we arrive at an upper limit of 11 \kms\ for the turbulent velocity. The validity of these assumptions are discussed further in \S \ref{sect:discussmass}. Now we assume a turbulent velocity of 5 \kms\ for the following results.}}

The optical depth is found by multiplying the absorption cross-section per atom by the column density  ($N_2$ - the total number density of hydrogen atoms in the $n=2$ level per square meter):
$$\tau_{\nu}=\sigma_{\nu}N_{2}$$

Therefore the EW is found for each line by integrating over the line profile:

$$EW = \int{\left(1-\exp^{-\tau_{\nu}}\right){d\nu}}$$ 

By considering a range of different values for the column density of atoms in the $n=2$ level we obtain a {\it{curve of growth}} for each line.  This describes how the EW of the line changes as we increase the density of absorbers.  At low densities (optically thin) the curve of growth is a linear increase in the EW proportional to the probability of the transition.  As we increase the column density eventually each line starts to saturate in its core leading to a slowing in the growth of the EW.  When the Doppler core is totally saturated (optically thick) then further growth of the EW occurs at a rate determined by the opacity in the Lorentzian line wings.

Once we have the EW of each line as a function of column density it is then possible to express each line's EW as a fraction of EW($H_{\alpha}$). These ratios are shown as the solid lines in the left panel of Fig.\ref{fig:hcol}. This shows that \ha\ starts to saturate first. Its growth of EW slows and the relative strength of the later lines increases. One by one the other lines start to saturate also.  Therefore by measuring the ratio of the observed EWs from the timeseries it should be possible to find a unique value for the column density of each prominence.

\subsubsection{Measuring the EW ratios}

As mentioned earlier it is unfortunately impossible to use \he\ in this analysis due to it being blended with the CaII H line. Even those parts of the line not directly superimposed on the calcium line could be subject to off-disc emission from it (as discussed in \S \ref{sect:emiss}).  We also had to abandon measuring the \height\ line due to anomalous results caused by a very strong neighbouring metal line.  The star-spot residuals in this line appeared strong enough to affect our prominence EWs.  Thus the analysis is now restricted to: \ha, \hb, \hg, \hd, \hnine\ and \hten.

\begin{figure*}
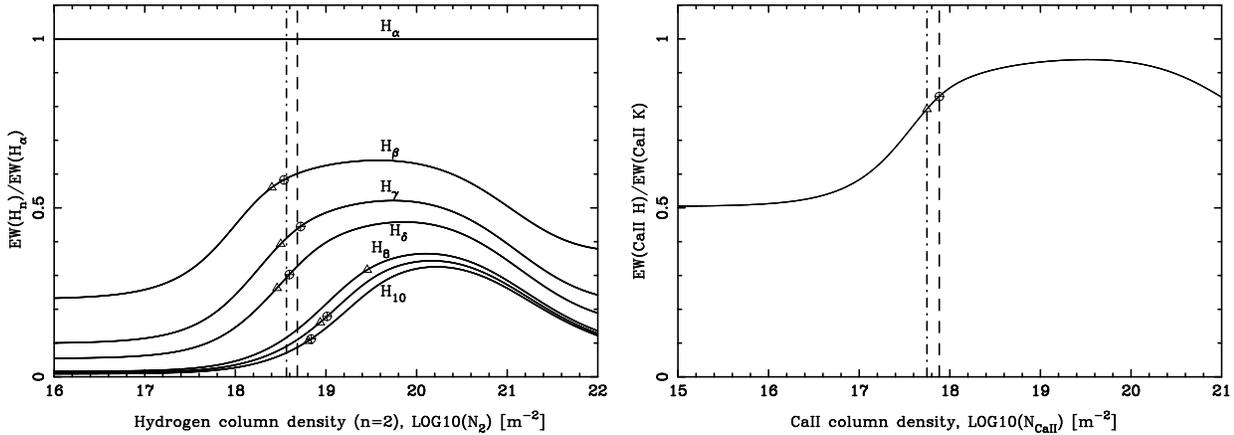

  \begin{tabular}{cc}
    \includegraphics[height=8cm,angle=270]{Image/njd_speedy06_fig5.ps} &
    \hspace{-2mm}
    \includegraphics[height=8cm,angle=270]{Image/njd_speedy06_fig6.ps} \\
  \end{tabular}
\caption[]{Obtaining column densities in n=2 of hydrogen (left panel) and CaII (right panel).  Solid lines show theoretical curve of growth EW ratios, as computed in \S \ref{sect:curves} for an example turbulent velocity of 5 \kms.  Circles show intersection with observed ratios for prominence A and triangles for prominence B.  The dashed vertical line is the optimal average column density for prominence A and the dot-dashed line for prominence B.  Note \he\ is absent due to blending with CaII H line and that \height\ is anomalous (see text).}
\protect\label{fig:hcol}
\end{figure*}

To measure the EW of the prominence absorption signatures it is first necessary to remove the `time-invariant' background stellar spectrum. For \speedy\ this is a rather more difficult task that one might first anticipate, the problem being that much of the timeseries is in strong absorption due to the prominence material. Other parts of the timeseries, however, show strong surface emission features (plage regions). {{The nature of the background surface spectrum for the regions covered by large prominences is unknown from \ha\ alone. It could well be that these phases are in even stronger emission. By examining the later Balmer lines in Fig. \ref{fig:hall}, where even the strongest prominences become optically thin, we can see that this is probably not the case. It is also important to remember that the bright surface plage regions contribute to the flux that the prominences receive and so should not simply be ignored. We finally chose to represent the background spectrum by taking the mean of the `light' half of the timeseries which is devoid of strong prominences.  Specifically the phases 0.65 to 1.0 were used to create the mean profile that was then subtracted from the entire timeseries. Initially we tried a much smaller range of phases which apparently had no prominences. However this produced an asymmetric mean profile and upon further examination a bright surface feature was present. Using this mean profile we obtained column densities that differed by a factor of 1.25 to 2.5 from those reported below. This is similar to the level of uncertainty we find on the prominence column densities.}}

We use the high S/N \ha\ timeseries to identify which pixels in the timeseries are to be used for the analysis.  In practice this was done by selecting a box around a single prominence feature and selecting all the negative pixels within that box.  The EWs are then simply the ratio of the sum of the corresponding pixels in each line to that of \ha. Associated uncertainties on these ratios are also derived.

The points where the observed EW ratios intersect the theoretical ratio of the curves of growth are shown for each line in Fig. \ref{fig:hcol}.  The uncertainties in measured EWs are then translated into uncertainties in column density using the model curves of growth ratios.  Thus we can take an {{inverse variance weighted average}} of these points to obtain a single estimate for the column density in the n=2 level of hydrogen.  The final uncertainty on the column density of each prominence is estimated from the standard deviation of the different line ratios.

Obviously the high S/N of the stronger Balmer lines like \hb\ allow us to measure the ratio of their EWs to \ha\ to a much greater precision than some of the weaker, later, Balmer lines.  However because it turns out that the cloud is quite optically thick in \hb\ it means that there is a larger uncertainty on the column density once the measurement uncertainty has been converted to an uncertainty in the column density.  Therefore \hg\ and \hd\ have most weight when calculating the optimal column density.

This technique only works for the very strong prominence features.  As already discussed there are many weaker prominences that cross the star's brighter hemisphere (phases 0.6 - 1.0), most of which can only be detected in \ha\ and \hb.  We therefore limit our analysis to the three strongest prominence signatures on August 02 (labelled A,B and C in Fig.\ref{fig:rtseries}) and the strongest prominence on August 07 (labelled D in Fig.\ref{fig:rtseries}).

\begin{table*}
\caption{The measured values and derived quantities are display along with their uncertainties. These results assume a turbulent velocity of 5 \kms\ (see text). Note that there is a systematic uncertainty on the CaII column density of log $N_{CaII}$=$\pm$0.25 (see text). We also report the average and standard deviation of the four prominences for various physical properties.}
\protect\label{tab:results}
\begin{center}
\begin{tabular}{ccccccccccccc}
\hline
\multicolumn{1}{c}{\it{Prominence}}	& \multicolumn{1}{c}{{\it{Fractional}}} & \multicolumn{2}{c}{\it{Optical depths}} & \multicolumn{3}{c}{{\it{Column densities}} [m$^{-2}$]}	& \multicolumn{1}{c}{\it{Areas}} & \multicolumn{1}{c}{\it{Masses}}	\\
& {\it{Depth}} & $\tau$(H$_{\alpha}$) & $\tau$(CaII K) & log $N_{2}$ &  log $N_{CaII}$   & log $N_{1}$ & log $A$ [$\mathrm{m^{2}}$] &  log $M$ [$\mathrm{kg}$] \\

\hline	
A & 0.12$\pm{0.015}$ & 25.7 & 21.3 & 18.68$\pm{0.19}$ & 17.88$\pm{0.25}$ &  23.53$\pm{0.25}$ & 17.31$\pm{0.06}$ & 14.06$\pm{0.26}$ \\
B & 0.25$\pm{0.005}$ & 19.5 & 15.8 & 18.56$\pm{0.21}$ & 17.75$\pm{0.25}$ &  23.40$\pm{0.25}$ & 17.63$\pm{0.03}$ & 14.25$\pm{0.25}$ \\
C & 0.19$\pm{0.01}$ & 22.4 & 13.5 & 18.62$\pm{0.19}$ & 17.68$\pm{0.25}$ &  23.33$\pm{0.25}$ & 17.51$\pm{0.04}$ & 14.06$\pm{0.25}$ \\
D & 0.24$\pm{0.01}$ & 21.9 & 11.7 & 18.61$\pm{0.18}$ & 17.62$\pm{0.25}$ &  23.27$\pm{0.25}$ & 17.61$\pm{0.04}$ & 14.10$\pm{0.25}$ \\
\hline
{\it{Averages}} & & $22.4\pm2.6$ & $15.6\pm4.2$ & $18.61\pm0.05$ & $17.73\pm0.11$ & $23.38\pm0.11$ & $17.52\pm0.15$ & $14.12\pm0.09$\\
\hline
\end{tabular}
\end{center}
\end{table*}

The results of this analysis are shown in Table \ref{tab:results}.  We find that the prominences have a relatively small spread of column densities in the $n=2$ level of hydrogen, with the mean being $\mathrm{log}\ {N}_{2}=18.61\pm0.05\ \mathrm{m^{-2}}$.  Prominence A has the highest column density at $\mathrm{log}\ {N}_{2}=18.68\pm0.19\ \mathrm{m^{-2}}$. As the uncertainty on each individual prominence is greater than the scatter in observed prominence column densities, we quote a final representative column density of $\mathrm{log}\ {N}_{2}=18.61\pm0.2\ \mathrm{m^{-2}}$.  Multiplying the column densities by the central absorption cross-section in each Balmer line we find the optical depths of prominences in each line.  In Table \ref{tab:results} we quote optical depths in \ha\ for each prominence.  These range between $19<\tau<26$ confirming that the prominences are indeed optically thick at the \ha\ line centre.  The optical depth drops to unity at around \hd\ or \height\ depending upon the individual prominence.

{{For the results shown in Table \ref{tab:results} we assumed a turbulent velocity of 5 \kms.  As this is considerably smaller than the thermal Doppler velocity of 12.9 \kms\ turbulence had little effect on the resulting column densities. If we increase the turbulent velocity to the upper limit of 11 \kms\ the column density increases by $\mathrm{log}\ {N}_{2}=0.1\ \mathrm{m^{-2}}$ (or a factor of ~1.25). This is smaller than the uncertainty quoted on each prominence column density in Table \ref{tab:results}.}}

\subsection{Column density of CaII}
\protect\label{sect:caii}
In order to find the masses of the prominences it is necessary to look at resonance lines, that is transitions down to the ground state.  While we were able to use many Balmer lines to estimate the column density in the n=2 level, the number of atoms in the ground state (n=1) is so far unknown.  The ratio between the populations of the n=1 to n=2 depends on the excitation of the cloud.  Two strong lines of singly ionised calcium, CaII H\&K, are transitions down to the ground state.

We plot the CaII H\&K timeseries spectra in Fig. \ref{fig:caii}.  Just as for the each of the Balmer lines in Fig. \ref{fig:hall}, we have subtracted a mean stellar spectrum obtained from the same range of phases.  The overall general appearance of the prominence pattern looks quite similar to that in the Balmer series.  One half of the timeseries shows dark prominence absorption signatures crossing the stellar disc.  The other half shows bright plage active regions on the stellar surface.  Also visible in the timeseries is a stellar flare at $\phi\approx0.3$. 

This CaII H\&K dataset has already been published by \cite{wolter05b} in the context of using Doppler imaging to map the \speedy\ chromosphere.  They refer to unexplained `changes on rapid timescales'.  As can be seen from Fig.\ref{fig:caii} these are almost certainly due to the prominence material transiting the stellar disc.

\subsubsection{{{Calculating the curve of growth}}}

We again use the ratio of the measured prominence EWs in order to ascertain the column density in CaII.  The curves of growth are calculated in the same manner as for hydrogen.  {{ As in \S \ref{sect:curves} the turbulent velocity is fixed at 5 \kms\ for the following work.}}

The oscillator strength of the CaII K line is exactly twice that of the CaII H line.  Therefore when examining optically thin material the EW of CaII H will be half that of CaII K.  As we increase the column density and the CaII K line starts to saturate, the rate of EW growth slows, so the ratio of the strengths of the two lines decreases. This is shown in the right-hand panel of Fig. \ref{fig:hcol}. 

\subsubsection{{{Measuring the EW ratios}}}
In order to measure the EWs both CaII lines were binned onto the same wavelength scale as the Balmer lines. The same points in the timeseries identified from the \ha\ timeseries were used to find the EWs.  The only difference, however, is that we only used the left hand side, the side with negative velocity with respect to the centre of both lines. This is because of the \he\ contamination of the CaII H line.  This extends to just under half of the CaII H line profile.  To investigate what effect this may have on our results we performed the same analysis on the hydrogen Balmer lines and compared it to our results in \S\ref{sect:hcol} which used the whole of the timeseries.  We found that this produced a scatter in the derived column densities of up to $\mathrm{log}\ {N}_{2}=0.2\ \mathrm{m^{-2}}$ (or a factor of ~1.6).  However the order of column densities remained unchanged, e.g. prominence A was still the most optically thick prominence.

\begin{figure}
\begin{center}
\includegraphics[height=8cm,angle=270]{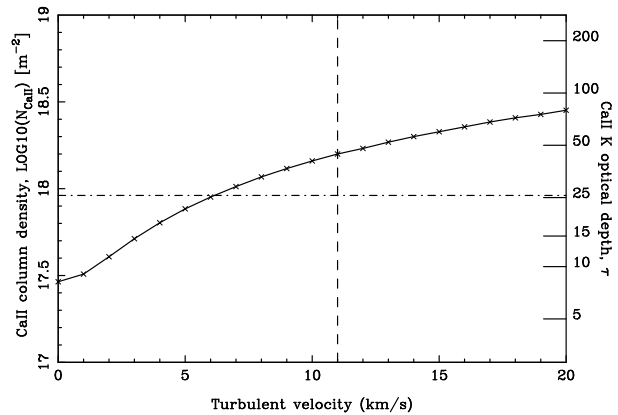}
\end{center}
\caption[]{{{The CaII column density is plotted as a function of the turbulent velocity for prominence A. The axis on the right of the plot shows the corresponding optical depth of the CaII K line.  The vertical dashed line at 11 \kms\ gives an observational upper limit on the turbulence.  The dot-dash horizontal line shows the optical depth of the \ha\ line for reference.}}}
\protect\label{fig:caiiturb}
\end{figure}

From Fig.\ref{fig:hcol} there appears to be an ambiguity on the CaII column density due to there being two potential solutions.  The results using the first (lower column density solution) are shown in Table \ref{tab:results}.  A similar spread is observed in the column densities of CaII as in the Balmer lines with a mean of ($\mathrm{log}\ {N}_{CaII}=17.73\pm0.11\ \mathrm{m^{-2}}$).  Prominence A is also found to have the highest column density at $\mathrm{log}\ {N}_{2}=17.88\ \mathrm{m^{-2}}$.  We also quote in Table \ref{tab:results} the optical depth of each prominence in CaII K, finding $12<\tau<21$, less than the optical depths in \ha\ ($19<\tau<26$). This is actually similar to the optical depths in \hb. Visual comparison of the \hb\ line in Fig.\ref{fig:hall} with the CaII K line in Fig.\ref{fig:caii} shows the prominence pattern to be similar in strength. This is why we can safely exclude the second solution shown on Fig.\ref{fig:hcol}.  The second solution gives a column density approximately a thousand times higher than that of the first solution.  Therefore the prominence optical depths in CaII K would also be a thousand times higher. Such optical depths are simply not supported by the visual appearance of the CaII K line in Fig.\ref{fig:caii} when compared to the hydrogen lines in Fig.\ref{fig:hall}. 

As there are only two calcium lines, the uncertainty in the derived CaII column densities are more difficult to estimate. The formal errors on each individual CaII column density are almost certainly underestimates of the true uncertainties.  The normalisation of the continuum in these blue orders is likely to be the primary source of error due to the difficulty of locating the continuum.  The CaII H\&K lines are in subsequent echelle orders. We ran a Monte Carlo simulation assuming that the continuum of each of these orders may be inaccurate by up to $\pm5\%$.  The standard deviation of the resulting probability distribution provides a systematic uncertainty on all calculated CaII column densities giving $\mathrm{log}\ {N}_{CaII}=17.73\pm0.25\ \mathrm{m^{-2}}$.  A further check was made using a template reference star taken during the observing run, GL 472.  This is a K0V dwarf observed as a slow rotating reference template for use in Doppler imaging.  The narrow lines of this star make the continuum easier to fit.  We therefore performed the same analysis, using instead the CaII H\&K continuum fits of GL 472.  The resulting average is $\mathrm{log}\ {N}_{CaII}=17.63\ \mathrm{m^{-2}}$, very similar to the result quoted above and within the systematic uncertainty determined above.

\begin{figure*}
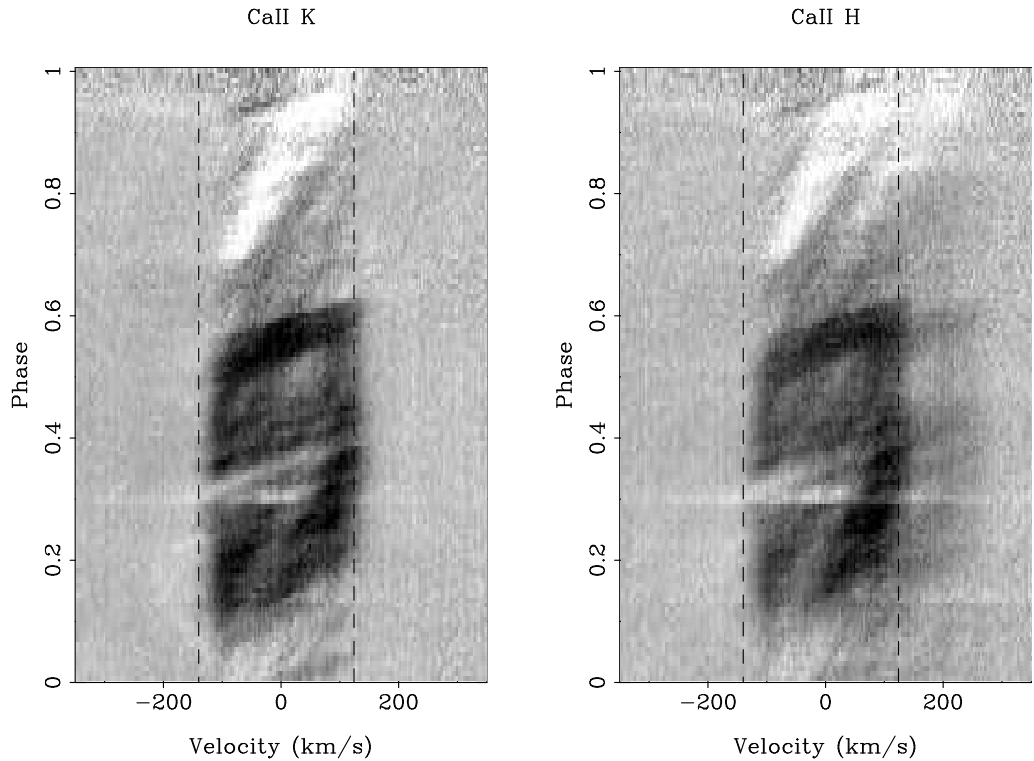

\begin{center}

  \begin{tabular}{cc}
    \includegraphics[height=10cm,angle=0]{Image/njd_speedy06_fig8.ps} &
    \hspace{4mm}
    \includegraphics[height=10cm,angle=0]{Image/njd_speedy06_fig9.ps} \\
  \end{tabular}
    
\end{center}
\caption[]{{{The August 02}} CaII K (left) \& H (right) timeseries are shown with contrast set so that black is at 0.85 and white is at 1.05 times the local continuum. An average profile has been subtracted (see text). {{Dashed vertical lines show the \vsini\ limits}}. Again, note that the CaII H line is blended with the \he\ line.  Bright surface plage regions are visible as well as weak off-disc emission.  A stellar flare is present at phase $\approx0.3$.}
\protect\label{fig:caii}
\end{figure*}

{{We investigate the result of increasing the turbulent velocity from 0 - 20 \kms\ in Fig. \ref{fig:caiiturb}. A vertical dashed line on the plot shows the maximum allowed turbulent velocity of 11 \kms. The resulting CaII column density increases by a factor of five over the range 0 - 11 km/s from $\mathrm{log}\ {N}_{CaII}=17.46\ \mathrm{m^{-2}}$ to $\mathrm{log}\ {N}_{CaII}=18.2\ \mathrm{m^{-2}}$.  This is discussed further in \S \ref{sect:discussmass}.}}

\subsection{Weighing prominences}
\protect\label{sect:weigh}

From the previous sections we now have the column densities and optical depths for the four prominences in the n=2 level of hydrogen and in the ground state of singly ionised calcium.  In order to calculate the mass of each prominence we need to know the column density in the ground state of hydrogen.

At the expected prominence temperatures of approximately 10,000 K we assume that the calcium atoms are singly ionised.  Then we can use solar abundance ratios to find the amount of neutral hydrogen. The ratio of calcium to hydrogen in solar prominences is approximately $\mathrm{log}\ \frac{H_{n=1}}{Ca}=5.65$ \citep{gouttebroze02}.  We adopt this ratio to obtain column densities in the ground state of hydrogen, displayed in Table \ref{tab:results}.  

We now require an estimate of the area of the prominences. Here we use the same approximation as \cite{cam1990}.  From \S\ref{sect:hcol} we know that the the prominences are optically thick at \ha\ ($\tau\approx22$) line centre.  Therefore when they transit the centre of the stellar disc, the fractional depth of the absorption profile gives us the fraction of the star obscured by the prominence.  This is analogous to how the fractional depth of a planetary transit gives the ratio of the areas of the planet and star. We use this to calculate the absolute areas of the prominences, using the \speedy\ radius of $1.06\pm{0.04}\ \mathrm{R}_{\odot}$ found in Paper I. These are shown in Table \ref{tab:results}.  The average area is found {{to be 20\% of the stellar disc.}} The masses of the prominences can now be calculated as simply the product of the hydrogen column density, the prominence area and the mass of a hydrogen atom:
$$M = m_{H}N_{1}A$$

The results are again shown in Table \ref{tab:results}. Prominences A, C and D are found to have very similar masses while prominence B has considerably higher mass. {{For a turbulent velocity of 5 \kms\ the average of the four prominences is $1.3 \mathrm{x}10^{14}\ \mathrm{kg}$.   We note that the individual uncertainty on each prominence arising from systematic error (as discussed in \S \ref{sect:caii}) exceeds the scatter in the prominence masses resulting in the range $0.7-2.3\ \mathrm{x}10^{14}\ \mathrm{kg}$.  

The CaII column densities found in the last section, from considering the full range of permitted turbulent velocities, provide an estimate of the minimum and maximum prominence masses.  These are found to be $0.5-3.4\ \mathrm{x}10^{14}\ \mathrm{kg}$.  We discuss these limits further in \S \ref{sect:discussmass}.}}

\section{Prominences seen in emission}
\protect\label{sect:emiss}

Solar prominences are visible as bright emission features when they are seen off the limb of the Sun.  In theory it should be possible to detect stellar prominences as bright emission features when they are off the limb of the star. The source function of a prominence will be stellar \ha\ photons that have been scattered (see \S \ref{sect:intro}) into our line-of-sight by the prominence, plus any additional source function from the thermal properties of the cloud itself.  The relative intensity we observe will be proportional to the size and optical thickness of the prominence in \ha\ and inversely proportional to the {{square of the prominence height}} above the stellar surface.  As \speedy\ prominences are found at least two stellar radii above the stellar surface these features are likely to be very faint.  It is also worth noting that, depending upon the geometry of the system, a prominence seen off the stellar disc need not transit the star as seen from our line of sight.

\subsection{Detecting loops of emission}

\begin{figure*}
\begin{center}

  \begin{tabular}{ccc}
    \includegraphics[height=9cm,angle=0]{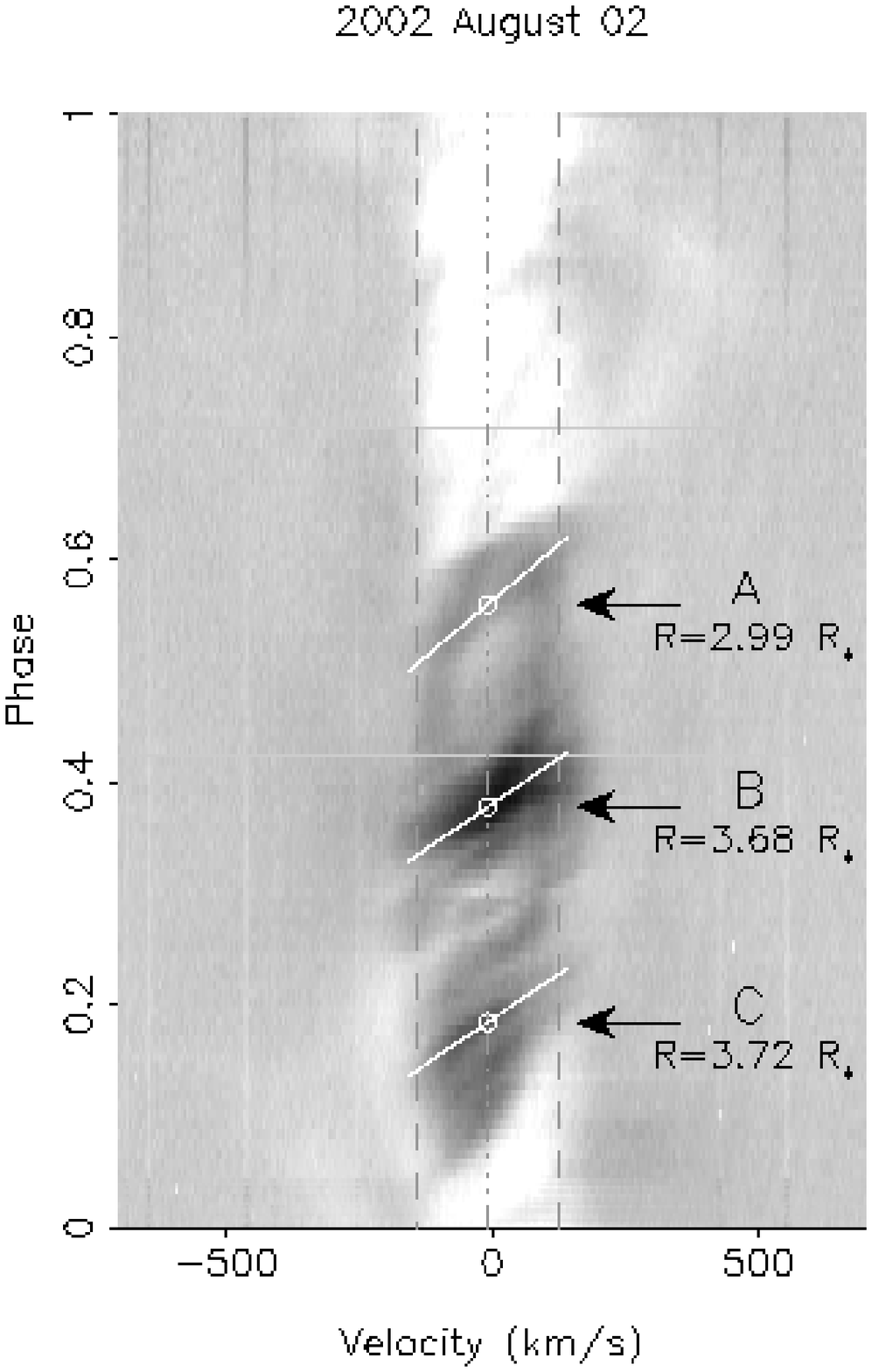} &

    \includegraphics[height=9cm,angle=0]{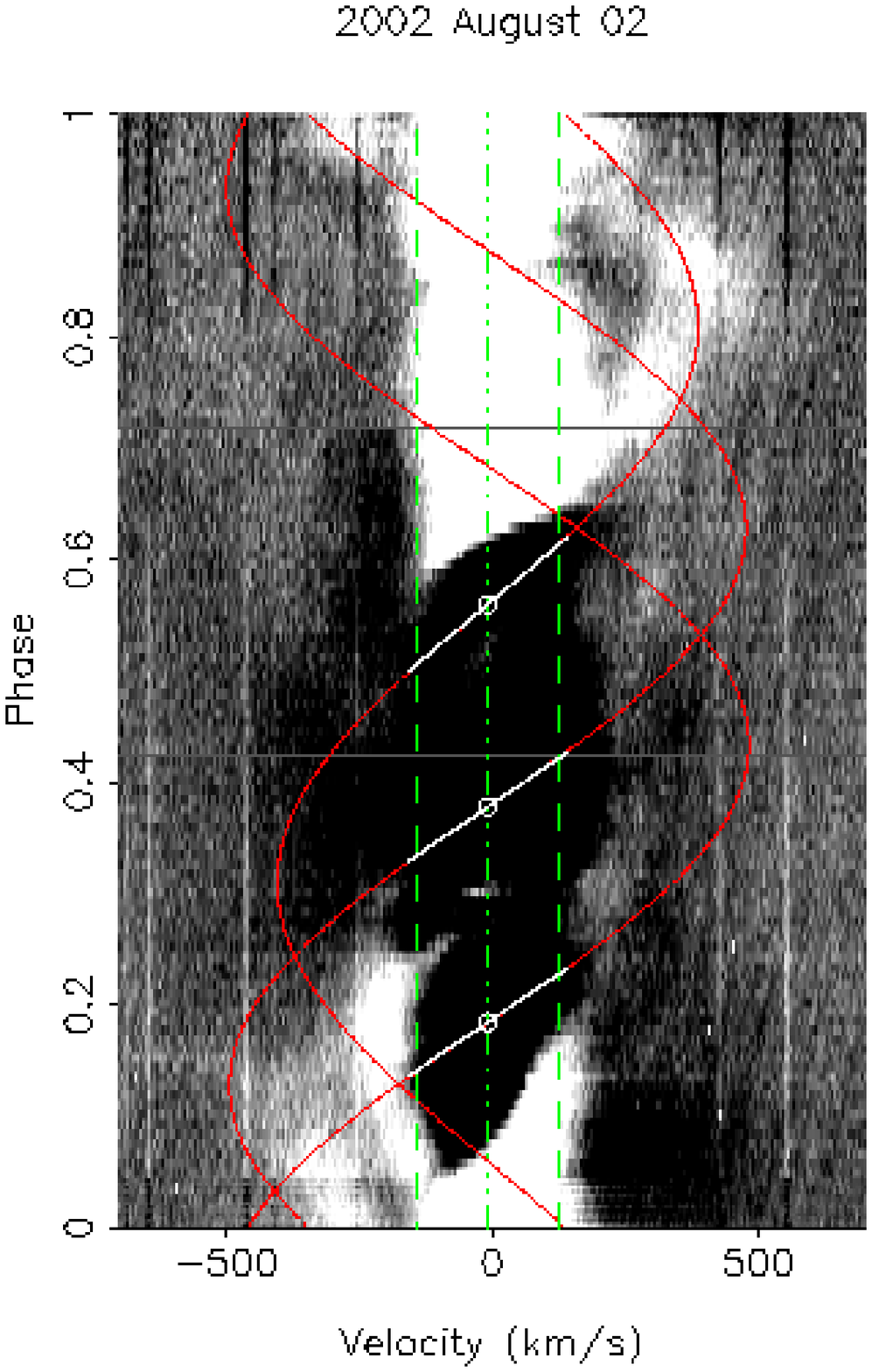} \\

    \includegraphics[height=9cm,angle=0]{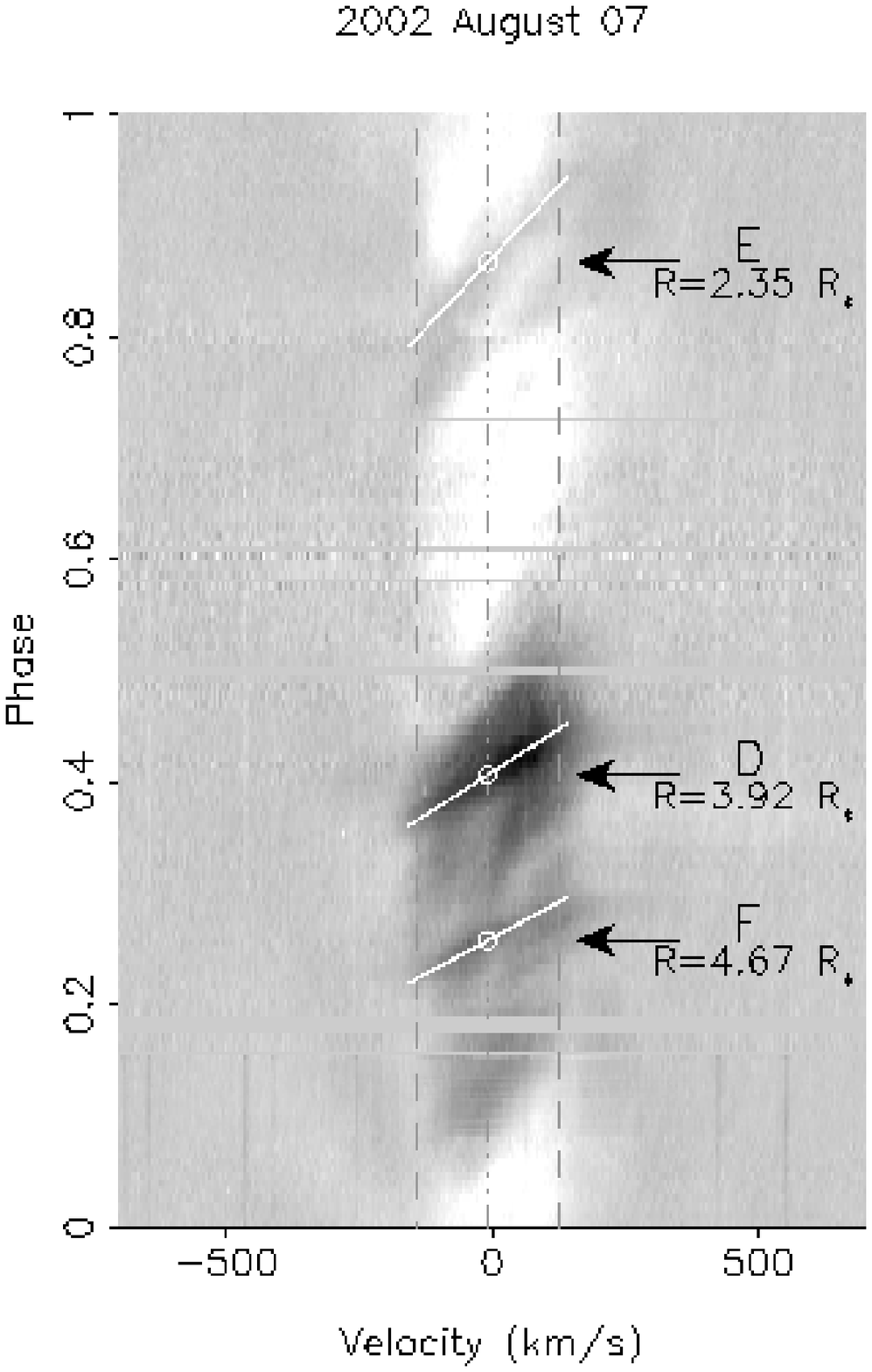} &

    \includegraphics[height=9cm,angle=0]{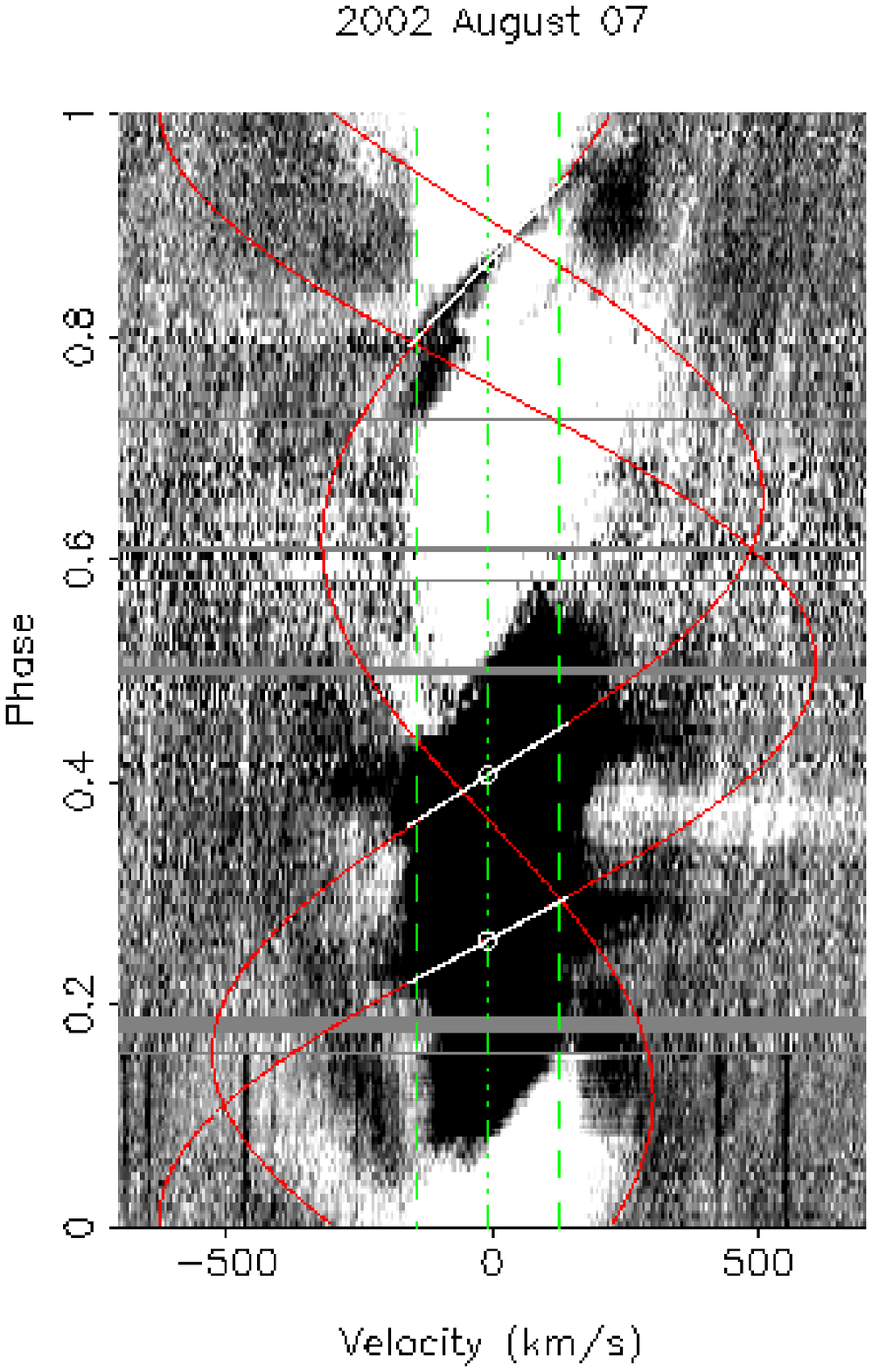} \\
    
  \end{tabular}

\end{center}
\caption[Off-disc emission]{Timeseries to show loops of off-disc emission.  The left panels show the fits to the transiting absorption features and have a grey-scale that runs from black at 0.85 to white at 1.05 times the local continuum level.  Similarly the top right panel is set 0.98 to 1.02, while the bottom right panel has 0.99 to 1.01, in order to highlight the weak emission. The fits from the left panels are extended as velocity sinusoids (red) on the right panels. Dashed vertical lines (green) show the \vsini\ limits.}
\protect\label{fig:emisscomb}
\end{figure*}

The exceptional signal-to-noise of the VLT \ha\ timeseries allows us to examine the regions beyond the stellar $v{\rm{sin}}i$ limits for evidence of off-disc emission.  {{To do this we created a mean profile, this time using the whole of the timeseries as there are no dominant strong features present outside of the stellar \ha\ profile. This was then subtracted from the entire timeseries to search for time-varying features.}}

We first need to ascertain a conservative estimate for our sensitivity to weak off-disc features.  The formal errors on each pixel, derived from Poisson statistics during the reduction, are underestimates as they take no account of systematic variations throughout the night.  Thus a featureless section of the timeseries adjacent to, but unaffected by, the \ha\ line and also avoiding any strong iron lines was chosen.  The standard deviation from the mean in this section which included all phases was $\sigma=0.004$, or 0.4\% of the local continuum level. This gives us a guide to the significance of any of the features we identify.

In the right-hand panels of Fig.\ref{fig:emisscomb} we show the resulting \ha\ timeseries for both August 02 and 07, with contrast limits set to reveal small fluctuations above or below the continuum level. Fig.\ref{fig:emisscomb} shows a number of bright emission loops. In particular there is a very distinct loop of emission on August 02 to the red (positive velocity) side of the \ha\ line, centred on $\phi\approx0.8$.  The emission peaks at 2.5\% of the continuum level but weakens beyond the maximum elongation.  The loop starts just after prominence A at $\phi\approx0.56$ finishes transiting the stellar disc.  One immediate observation is that there is no corresponding emission loop on the blue side of the \ha\ line.

There are two other loops of emission both centred on $\phi\approx0.15$.  The smaller loop shows very strong emission, peaking at 3\% of the continuum, but it cannot be separated from the \ha\ line. This is probably a prominence very close to the stellar surface. The larger loop peaks at around 2\% of the continuum level. Again it is apparent that there is no corresponding loop on the red side of the \ha\ line. The observations on August 07 are generally of a lower quality yet similar emission loops can be seen in Fig. \ref{fig:emisscomb}.

\subsection{Understanding the emission}

In order to attempt to understand the nature of the relationship between the on-disc absorption and off-disc emission features we can use what we have learnt from tracking the prominences in absorption. For a detailed explanation of this matched filter technique we refer the reader to Paper I. We use the calculated phases and heights displayed in Table \ref{tab:heights} to complete velocity sinusoids which we trace out in Fig.\ref{fig:emisscomb}. We do this for the three strongest prominences on the stellar disc for both nights.  These are prominences A, B and C on August 02 and prominence D on August 07. We also include the next two strongest prominences on August 07, labelled E and F.

\begin{table}
\caption{{{The results of the prominence tracking analysis are displayed with phases of meridian crossing and calculated heights.}}}
\protect\label{tab:heights}
\begin{center}
\begin{tabular}{ccc}
\hline
Prominence & Phase & Height ($\frac{{\varpi}}{R_*}$) \\

\hline	
A		& 0.560 	& 2.99$\pm$0.07 \\
B		& 0.378 	& 3.68$\pm$0.26  \\
C		& 0.184 	& 3.72$\pm$0.20  \\
D		& 0.406 	& 3.92$\pm$0.21  \\
E		& 0.867 	& 2.35$\pm$0.07 \\
F		& 0.258 	& 4.67$\pm$0.16  \\
\hline
\end{tabular}
\end{center}
\end{table}

In general there is good agreement between the extended sinusoids and the observed loops of emission. For example prominences A and B on August 02 clearly account for the two strong loops of emission. Prominence C, however, does not seem to correlate with any emission feature. A similar off-disc emission pattern is seen on August 07. Prominence D appears to account for most of the emission on both the blue and the red side of the timeseries. While the low lying prominence E appears to correlate with a loop of emission seen to the blue side of the line at $\phi\approx0.65$.  The high prominence F (like C) does not appear to be associated with any emission loop.

Furthermore, the paths of prominences as estimated from tracking the absorption features appear to have systematically greater velocity amplitudes than the majority of the corresponding off-disc emission features.  This can be explained in the context of the prominence optical depths as obtained in \S\ref{sect:masses}.  The average optical depth of these prominences was found to be $\tau=22.4$.  Thus as stellar photons travel through the prominence interior they are scattered into random directions, some into our line of sight.  However because of this scattering the beam loses intensity and so the number of photons reaching the highest parts of the prominence is reduced.  

\begin{figure}
\begin{center}
\includegraphics[height=8cm,angle=270]{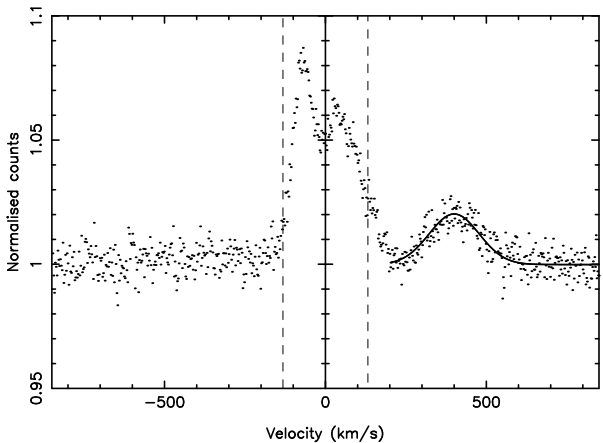}
\end{center}
\caption[]{{{An example of a single \ha\ spectrum at maximum elongation of the strongest observed loop of emission seen at $\phi\approx0.8$ on August 02.  The dashed lines show the stellar $v{\rm{sin}}i$ limits. The solid line shows the Gaussian model we have fitted to the emission, which is found to peak at $2\%$ above the continuum level. The FWHM is 260 \kms\ giving an estimate of the prominence thickness of 2.0 $\mathrm{R_{*}}$.}}}
\protect\label{fig:promthick}
\end{figure}

When finding prominence heights from tracking the absorption features we are naturally following the `lack of photons'.  Yet when we examine the off-disc emission we observe where these photons have gone.  As such, the observed emission provides us with information on the radial extent (or thickness) of the prominences. Clearly from a visual examination of Fig.\ref{fig:emisscomb} the prominences we observe do not extend to the stellar surface.  {{We scale a Gaussian model to the most prominent loop of emission on August 02, fitting for the width of the feature. An example fit for a spectrum taken at maximum elongation of this emission loop is shown in Fig.\ref{fig:promthick}. Both the amplitude and width of these emission features vary considerably along the emission loop.  We find a range of full width at half maximum (FWHM) that corresponds to a prominence thickness of between 1.7 and 2.5 $\mathrm{R_{*}}$.}}

\section{Discussion}
\protect\label{sect:discuss}

\subsection{Prominence stability}

In Paper I we investigated prominence stability on the timescale of days we found that many prominences could be re-observed on consecutive days, while a few were still recoverable some five nights later.  In \S \ref{sect:evol} we extended the timebase by comparing Paper I observations with the current data taken more than two weeks later.

The main result of this comparison is that the overall circumstellar distribution of prominence material remains the same. While there is clearly
evolution of the detailed structure of the individual prominences, the locations at which prominences are likely to be found are largely unchanged.
In particular prominence A was the structure we could most easily identify between the two epochs of observations.  We found that its phase of
observation, height and strength were all remarkably similar. This does not necessarily mean that this feature has remained unchanged for two weeks
(without observations of the intervening period it would be impossible to tell). What it does suggest however is that the magnetic structure within
which this prominence was trapped is stable on these timescales. Within this structure, individual prominences may have formed and either been
ejected or drained back to the surface within the two week gap when we do not have observations, but if this is the case, then another prominence
must have reformed in the same place and under very much the same conditions. 

It is interesting that this prominence, which appears to be the most stable, is also one of the largest. This is also true of solar prominences
\citep{schrijverbook01}. Small solar prominences which form close to  active regions tend to be most unstable while the longest-lived are also
typically the largest. In the solar case, the largest prominences are sheared out over their lifetimes by the action of differential rotation and
they are carried  poleward over a lifetime of months to form the polar crown. Their poleward migration is halted when they encounter the open field
of the polar cap. Whether something similar is happening on Speedy Mic is not clear.  Certainly, the apparent lack of high-latitude prominences
(discussed in \S \ref{sect:discussemiss}) suggests that the prominences are constrained to a range of latitudes, rather like their solar
counterparts. This is in stark contrast to the distribution of spots on Speedy Mic which can be seen at all latitudes, including the pole
(\citealt{barnes05}, {{\citealt{wolter05a}}}).

\subsection{Prominence masses}
\protect\label{sect:discussmass}

In \S\ref{sect:masses} we used the ratio of the strengths of multiple Balmer lines to determine the column densities of prominences in the first excited state of hydrogen to be ($\mathrm{log}\ {N}_{2}=18.61\pm0.2\ \mathrm{m^{-2}}$).  This is in good agreement with \cite{cam1990} who found ($\mathrm{log}\ {N}_{2}=18.4\pm0.2\ \mathrm{m^{-2}}$).  Consequently similar optical depths are also derived. From direct measurements of the EWs of prominences in the CaII H\&K lines, and then use of solar prominence abundance ratios, we were able to estimate the column density in the ground state of hydrogen.  This was found to be ($\mathrm{log}\ {N}_{1}=23.38\pm0.25\ \mathrm{m^{-2}}$).  Then using the projected areas of the prominences we estimated their masses to be $0.7-2.3\ \mathrm{x}10^{14}\ \mathrm{kg}$.  

{{The results outlined above were for a turbulent velocity of 5 \kms. The turbulent velocity could be in the range 0 - 11 \kms\ producing a larger prominence mass range of $0.5-3.4\ \mathrm{x}10^{14}\ \mathrm{kg}$. However there are reasons why we believe that the turbulent velocity probably lies towards the lower end of this range. Firstly we can examine the assumption that the smallest prominences are unresolved. The smallest features for which we could determine the widths have fractional depths in \ha\ that correspond to areas that cover at least 2\% of the stellar disc.  Assuming a circular cloud cross-section this gives a radius of 14 \% that of the \speedy\ radius.  Hence in velocity space a width of 18 km/s would be observed.  This in itself accounts for the observed widths. 

Another argument for small turbulent velocities comes from comparing the optical depths of prominences in the \ha\ and CaII K lines. On Fig. \ref{fig:caiiturb} we include a second axis that shows the corresponding optical depth in the CaII K line. For the range of turbulent velocities 0 - 11 \kms\ optical depths of $8<\tau<40$ are found.  We also plot on Fig. \ref{fig:caiiturb} a horizontal dot-dash line to show the optical depth at the centre of the \ha\ line at $\tau=25.7$. If we now compare the CaII K line in Fig. \ref{fig:caii} with the that of the hydrogen lines in Fig. \ref{fig:hall}, the prominence absorption features in the CaII K line appear to be less optically thick than the \ha\ line. This includes the lack of an emission loop associated with prominence A.  The CaII K timeseries is of sufficient S/N that we would expect to detect scattered photons from this loop if it had a similar optical depth to that of \ha. We suggest then that the optical depth of the CaII K line is less than the \ha\ line and thus the turbulent velocity should be less than 6 \kms. This is therefore similar to the level of turbulence seen in solar prominences. Hence our adopted turbulent velocity of 5 \kms\ seems valid in the previous sections. Therefore we arrive at a narrower range for prominence masses of ($0.5-2.3\ \mathrm{x}10^{14}\ \mathrm{kg}$).}}

We can now compare some of the physical properties of the \speedy\ prominence system found in \S\ref{sect:masses} with those of the only other similar analysis on a stellar prominence system.  Using different techniques to this work, \cite{cam1990} derived parameters for another rapidly rotating star, AB Dor. They found prominence masses of $4-5\mathrm{x}10^{14}\  \mathrm{kg}$, slightly greater than we find for \speedy.  Given the good agreement between this work and \cite{cam1990} on the column density of the n=2 level of hydrogen it seems unlikely that the above discrepancy on prominence masses represents a physical difference between AB Dor and \speedy\ prominences.  Rather it is probably just a difference between the two techniques employed due to the quality of data available at the time.  With the advent of new highly-efficient spectrographs and increased detector size and efficiency (particularly in the blue part of the spectrum) it is now possible to obtain simultaneous high-resolution spectra in the majority of the optical wavelength range.  In particular we are now able to obtain {\it{simultaneous}} observations of the \ha\ line and both CaII H\&K lines.  This was not possible at the time of the \cite{cam1990} observations, they therefore had to rely on a geometrical argument to obtain estimates of the column densities.

We can also compare \speedy\ prominences with their solar namesakes. For example a giant eruptive solar prominence was observed by \cite{gopalswamy98} to have an initial mass before eruption of $6\mathrm{x}10^{13}\ \mathrm{kg}$. {{This is comparable to the lower end of our mass estimate}}. However to have three such prominences on the Sun at any one time would be unusual indeed.

As we have the column densities in the first two levels of hydrogen we can estimate the excitation temperature of the prominences by assuming that they are in local thermodynamic equilibrium (LTE).  We note that this {{can not strictly}} be the case as else the {{10,000 K prominences would appear as emission rather than absorption features whilst transiting the stellar disc.}}  However given that we have shown \ha\ to be just optically thick, we consequently know that Lyman alpha must also be very optically thick.  As it is this transition that is largely responsible for detailed balance between the first two levels of hydrogen the LTE approximation should be reasonably accurate.

Comparing the column densities of hydrogen in the ground state with that in the first excited state, the average shows that there are {{$10^{4.77}=59,000$ times more atoms in the ground state than in the first excited state (assuming a turbulent velocity of 5 \kms)}}.  Therefore the ratio of the LTE level populations is given by the Boltzmann equation:
$$\frac{N_2}{N_1}=\frac{g_{2}}{g_{1}}\mathrm{exp}\left({-\frac{{\Delta}E}{kT}}\right)$$
where $g_{1}$ and $g_{2}$ are the statistical weights of the two levels, ${\Delta}E$ is the energy difference between them, $k$ is the Boltzmann constant and $T$ is the temperature sought.  This results in an excitation temperature of 9569 K.  This is very similar to the kinetic temperature of 10,000 K we used in section \S\ref{sect:masses}. While we realise that this is a circular argument, the results of this analysis are not found to depend strongly on the thermal Doppler temperature used in creating the theoretical line profiles.  {{For example if we start with a lower thermal temperature of 8000 K we obtain an excitation temperature of 9463 K.}}

\subsection{Distribution of prominence material}
\protect\label{sect:discussemiss}

In \S \ref{sect:emiss} we showed that some of the strongest absorption features (the largest prominences) could be observed as loops of emission.   However we found no emission counterparts for prominences C and F.  This probably reflects the relationship between the size of the prominence and the height above the stellar surface.  Here C and F cover slightly larger areas of the stellar disc than say prominence E.  Yet because prominence E is very much closer to the stellar surface ($R=2.35\ \mathrm{R_{*}}$) than either C or F ($R=3.72\ \mathrm{R_{*}}$ and $R=4.67\ \mathrm{R_{*}}$ respectively), and the intensity drops as the square of the distance, it is seen in emission whilst prominences C and F are not. 

If prominences were evenly distributed in latitude around \speedy\ then one would expect to observe loops of emission which can not be associated with prominences crossing the stellar disc.  However this seems not to be the case, from Fig. \ref{fig:emisscomb} we {\em{can}} associate all the observed emission loops with transiting prominences.  If large prominences, at similar heights to those observed, existed at latitudes of around $60\degr$ then they should be visible as emission loops.  So this either suggests that \speedy\ does not have such large high-latitude prominences, or that they must be extremely centrifugally flattened {{towards}} the equatorial plane. {{It should be noted though that if higher latitude prominences were smaller or indeed higher then they may exist without detectable emission.}}

Similar conclusions have been reached before from the lack of prominences observed on low-inclination rapid rotators.  \cite{jeffries94} found no prominences {{transiting the stellar disc}} around the rapidly rotating K dwarf BD+22$\degr$4409 which is observed at an inclination of $50\pm10\degr$.  {{Unfortunately, as discussed in the \S \ref{sect:intro}, the \speedy\ inclination is relatively poorly constrained and restricts what we can determine.  If the inclination is actually the $70\degr$ that both Doppler imaging teams found then even a prominence above $R=3\ \mathrm{R_{*}}$, located in the equatorial, plane would not be seen crossing the stellar disc at all. Indeed, these observations may favour a higher stellar inclination.}}

Speedy Mic is not the first rapidly rotating young star to show rotationally modulated emission outside of the \ha\ line.  Two G dwarfs RXJ1508.6-4423 \citep{donati00rx} and AP149 \citep{barnesap149} also show emission beyond their stellar \vsini\ limits which were attributed to a ring (or torus) of prominence material.  These two stars are very different cases to \speedy\ though. On neither star are prominences seen transiting the stellar disc, {{nor is there any evidence that some prominences eclipse the off-disc emission of others}}.  This is probably as both stars are seen at low inclination angles ($i=35\degr$ and $i=30\degr$ respectively).  Furthermore the strength of the emission is very different.  The amplitude of the emission in these systems is around 20\% of the local continuum as opposed to the 2-3\% we observe on \speedy.  This is probably due in part to the increased ionising UV flux from these early-type G dwarfs.  This coupled to the fact that the co-rotation radius on these stars is closer to the surface than in \speedy\ means the prominences receive greater irradiation. This drives the \ha\ source function closer to the thermal equilibrium value than to the pure-scattering value. 

Our analysis of the \speedy\ prominence system has considered only individual prominences that we can track across the stellar disc.  This is a simplification of the real distribution of prominence material.  For example if we look back at the \ha\ timeseries shown in Fig.\ref{fig:hall} we can see that prominences A and B are not totally separate entities.  The \ha\ line is still in weak absorption at the phases between these two prominences.  Yet if we compare this with the later Balmer lines (like \hg\ in the panel below \ha\ in Fig.\ref{fig:hall}) the absorption weakens considerably. This indicates that this material has a lower column density than the material in what we refer to as prominences A and B.  The large prominences that we have been studying therefore appear to be more like dense condensations in a `semi-torus' of gas.  {{This would also help to explain why some prominences do not show off-disc emission loops on {\em{both}} sides of the \ha\ line, as observed in \S \ref{sect:emiss}.  It is the intervening material between some of the prominences that obscure us from observing this emission, i.e. the prominence system is `self-eclipsing'.}}  Therefore there appear to be similarities between the \speedy\ prominence system and that of the low inclination systems RXJ1508.6-4423 and AP149.  The observed differences between these systems and Speedy Mic in many respects appear to be due to the different stellar inclinations.

\section{Conclusions}
\protect\label{sect:conc}

The first analysis of the evolution of a stellar prominence system on a two to three week timescale has been performed.  Whilst we see considerable evolution of most individual prominences the overall prominence pattern remains broadly similar.  One half of the stellar rotation is still covered in stronger prominence signatures than the other. We have shown that at least one individual prominence supporting structure appears to be stable over two weeks.  Any theoretical models for prominence support will have to be able to reproduce such stability.

We show that the prominence pattern is visible in the first eight lines of the hydrogen Balmer series and the CaII H\&K lines.  The column density of hydrogen in the first excited state and the derived \ha\ optical depth are very similar to that of the prominences on AB Dor reported by \cite{cam1990}.  This analysis, however, finds a lower total hydrogen column density from direct observations of CaII.  {{This leads us to find prominence masses of $0.5-2.3\ \mathrm{x}10^{14}\ \mathrm{kg}$, intermediate between the masses of the largest solar prominences and those of AB Dor. We suspect that the difference in the derived masses of the prominence systems on these two rapidly rotating stars is not physical but is probably a result of the different techniques used and the data quality available.}}

We have identified loops of emission outside of the rotationally broadened \ha\ profile.  These are shown to be caused by the same prominences we see transiting the stellar disc in absorption.   The emission loops confirm the heights found from tracking the absorption signatures as the visible emission is shown to appear at smaller radii in all cases.  This observation can be explained by the optical thickness of the clouds in \ha.  By combining these two methods of observing prominences we have been able to develop a more complete picture of the geometry and physical parameters of the prominence system.  All the strong loops of emission we observe can be associated with particular prominences seen transiting the stellar disc.  This limits the number of large high-latitude prominences.  Furthermore, combined with the high degree of self-eclipse observed in the \speedy\ prominence system we suggest that it is indeed highly flattened, most probably near to the stellar equatorial plane.  This apparent flattening of the prominence system could result from the presence of a unidirectional field in the polar cap. If there is a strong dipolar component in the stellar field then prominences will form in the equatorial plane \citep{mcivor03}.  From Doppler imaging of the \speedy\ surface \cite{barnes05} found that there was no single uniform polar spot. So it may be the case that the stellar field has a high degree of complexity and prominences that form at high latitudes are centrifugally driven into the equatorial plane.

It is clear that the \speedy\ prominence system is complex, with the distribution of co-rotating prominence material not restricted to the dense clumps that we have identified as prominences.  We note that there seems to be interesting parallels between these observations and those of the low inclination G dwarfs RXJ1508.6-4423 and AP149.  In the future it may be informative to develop a model where we could change the physical parameters and the viewing angle of such prominence systems. The results could then be compared directly with these observations.

\section{ACKNOWLEDGEMENTS}

Based on observations made with the European Southern Observatory telescopes obtained from the ESO Science Archive Facility.  The data in this paper were reduced using {\sc starlink} software packages.  NJD wishes to acknowledge the financial support of a UK PPARC studentship.  

\bibliographystyle{mn2e}
\bibliography{iau_journals,master,ownrefs,njd2}

\begin{thebibliography}{}

\bibitem[\protect\citeauthoryear{{Barnes}}{{Barnes}}{2005}]{barnes05}
{Barnes} J.~R.,  2005, MNRAS, 364, 137

\bibitem[\protect\citeauthoryear{{Barnes}, {Collier Cameron}, {James} \&
  {Steeghs}}{{Barnes} et~al.}{2001}]{barnesap149}
{Barnes} J.~R.,  {Collier Cameron} A.,  {James} D.~J.,    {Steeghs} D.,  2001,
  MNRAS, 326, 1057

\bibitem[\protect\citeauthoryear{{Cameron}, {Jardine}, {Wood} \&
  {Donati}}{{Cameron} et~al.}{2003}]{cam2003rev}
{Cameron} A.~C.,  {Jardine} M.,  {Wood} K.,    {Donati} J.-F.,  2003, in EAS
  Publications Series {Stellar prominences and coronal magnetic fields}.
pp 217--+

\bibitem[\protect\citeauthoryear{{Collier Cameron}, {Duncan}, {Ehrenfreund},
  {Foing}, {Kuntz}, {Penston}, {Robinson} \& {Soderblom}}{{Collier Cameron}
  et~al.}{1990}]{cam1990}
{Collier Cameron} A.,  {Duncan} D.~K.,  {Ehrenfreund} P.,  {Foing} B.~H.,
  {Kuntz} K.~D.,  {Penston} M.~V.,  {Robinson} R.~D.,    {Soderblom} D.~R.,
  1990, MNRAS, 247, 415

\bibitem[\protect\citeauthoryear{{Collier Cameron} \& {Robinson}}{{Collier
  Cameron} \& {Robinson}}{1989a}]{cam1989a}
{Collier Cameron} A.,  {Robinson} R.~D.,  1989a, MNRAS, 236, 57

\bibitem[\protect\citeauthoryear{{Collier Cameron} \& {Robinson}}{{Collier
  Cameron} \& {Robinson}}{1989b}]{cam1989b}
{Collier Cameron} A.,  {Robinson} R.~D.,  1989b, MNRAS, 238, 657

\bibitem[\protect\citeauthoryear{{Collier Cameron}, {Walter} \&
  {Vilhu}}{{Collier Cameron} et~al.}{1999}]{cam1999}
{Collier Cameron} A.,  {Walter} F.~M.,    {Vilhu} O.~e.,  1999, MNRAS, 308, 493

\bibitem[\protect\citeauthoryear{{Donati} \& {Collier Cameron}}{{Donati} \&
  {Collier Cameron}}{1997}]{donati97ab}
{Donati} J.-F.,  {Collier Cameron} A.,  1997, MNRAS, 291, 1

\bibitem[\protect\citeauthoryear{{Donati}, {Collier Cameron}, {Hussain} \&
  {Semel}}{{Donati} et~al.}{1999}]{donati99ab}
{Donati} J.-F.,  {Collier Cameron} A.,  {Hussain} G.~A.~J.,    {Semel} M.,
  1999, MNRAS, 302, 437

\bibitem[\protect\citeauthoryear{{Donati}, {Mengel}, {Carter}, {Marsden},
  {Collier Cameron} \& {Wichmann}}{{Donati} et~al.}{2000}]{donati00rx}
{Donati} J.-F.,  {Mengel} M.,  {Carter} B.~D.,  {Marsden} S.,  {Collier
  Cameron} A.,    {Wichmann} R.,  2000, MNRAS, 316, 699

\bibitem[\protect\citeauthoryear{{Dunstone}, {Barnes}, {Cameron} \&
  {Jardine}}{{Dunstone} et~al.}{2006}]{dunstone06}
{Dunstone} N.~J.,  {Barnes} J.~R.,  {Cameron} A.~C.,    {Jardine} M.,  2006,
  MNRAS, 365, 530

\bibitem[\protect\citeauthoryear{{Engvold}}{{Engvold}}{1998}]{engvold1998}
{Engvold} O.,  1998, in {Webb} D.~F.,  {Schmieder} B.,   {Rust} D.~M.,  eds,
  ASP Conf. Ser. 150: IAU Colloq. 167: New Perspectives on Solar Prominences
  {Observations of Filament Structure and Dynamics (Review)}.
pp 23--+

\bibitem[\protect\citeauthoryear{{Gopalswamy} \& {Hanaoka}}{{Gopalswamy} \&
  {Hanaoka}}{1998}]{gopalswamy98}
{Gopalswamy} N.,  {Hanaoka} Y.,  1998, ApJ, 498, L179+

\bibitem[\protect\citeauthoryear{{Gouttebroze} \& {Heinzel}}{{Gouttebroze} \&
  {Heinzel}}{2002}]{gouttebroze02}
{Gouttebroze} P.,  {Heinzel} P.,  2002, A\&A, 385, 273

\bibitem[\protect\citeauthoryear{Horne}{Horne}{1986}]{horne86extopt}
Horne K.~D.,  1986, PASP, 98, 609

\bibitem[\protect\citeauthoryear{Hussain, Brickhouse, Dupree, Jardine, van
  Ballegooijen, Collier~Cameron, Donati \& Favata}{Hussain
  et~al.}{2005}]{hussain_chandra1_05}
Hussain G.,  Brickhouse N.,  Dupree A.,  Jardine M.,  van Ballegooijen A.,
  Collier~Cameron A.,  Donati J.-F.,    Favata F.,  2005, ApJ, 621, 999

\bibitem[\protect\citeauthoryear{{Jardine}, {Collier Cameron} \&
  {Donati}}{{Jardine} et~al.}{2002}]{jardine02structure}
{Jardine} M.,  {Collier Cameron} A.,    {Donati} J.-F.,  2002, MNRAS, 333, 339

\bibitem[\protect\citeauthoryear{{Jardine} \& {van Ballegooijen}}{{Jardine} \&
  {van Ballegooijen}}{2005}]{jardine05}
{Jardine} M.,  {van Ballegooijen} A.~A.,  2005, MNRAS, 361, 1173

\bibitem[\protect\citeauthoryear{{Jardine}, {Wood}, {Collier Cameron}, {Donati}
  \& {Mackay}}{{Jardine} et~al.}{2002}]{jardine02xray}
{Jardine} M.,  {Wood} K.,  {Collier Cameron} A.,  {Donati} J.-F.,    {Mackay}
  D.~H.,  2002, MNRAS, 336, 1364

\bibitem[\protect\citeauthoryear{{Jeffries}, {Byrne}, {Doyle}, {Anders},
  {James} \& {Lanzafame}}{{Jeffries} et~al.}{1994}]{jeffries94}
{Jeffries} R.~D.,  {Byrne} P.~B.,  {Doyle} J.~G.,  {Anders} G.~J.,  {James}
  D.~J.,    {Lanzafame} A.~C.,  1994, MNRAS, 270, 153

\bibitem[\protect\citeauthoryear{{McIvor}, {Jardine}, {Cameron}, {Wood} \&
  {Donati}}{{McIvor} et~al.}{2003}]{mcivor03}
{McIvor} T.,  {Jardine} M.,  {Cameron} A.~C.,  {Wood} K.,    {Donati} J.-F.,
  2003, MNRAS, 345, 601

\bibitem[\protect\citeauthoryear{Mills}{Mills}{1994}]{mills92}
Mills D.,  1994, Technical Report~152, Starlink User Note.
Rutherford Appleton Laboratory

\bibitem[\protect\citeauthoryear{{Schrijver} \& {Zwaan}}{{Schrijver} \&
  {Zwaan}}{2000}]{schrijverbook01}
{Schrijver} C.~J.,  {Zwaan} C.,  2000, {Solar and Stellar Magnetic Activity}.
Cambridge University Press, New York

\bibitem[\protect\citeauthoryear{{Singh}, {Drake}, {Gotthelf} \&
  {White}}{{Singh} et~al.}{1999}]{singh99xray}
{Singh} K.~P.,  {Drake} S.~A.,  {Gotthelf} E.~V.,    {White} N.~E.,  1999, ApJ,
  512, 874

\bibitem[\protect\citeauthoryear{{Wolter} \& {Schmitt}}{{Wolter} \&
  {Schmitt}}{2005}]{wolter05b}
{Wolter} U.,  {Schmitt} J.~H.~M.~M.,  2005, A\&A, 435, L21

\bibitem[\protect\citeauthoryear{{Wolter}, {Schmitt} \& {van Wyk}}{{Wolter}
  et~al.}{2005}]{wolter05a}
{Wolter} U.,  {Schmitt} J.~H.~M.~M.,    {van Wyk} F.,  2005, A\&A, 435, 261

\end{thebibliography}

\end{document}